# Deep Learning based OTDOA Positioning for NB-IoT Communication Systems

Guangjin Pan, Tao Wang, Xiufeng Jiang, Shunqing Zhang
guangjin_pan@shu.edu.cn, twang@shu.edu.cn, XiufengJiang@shu.edu.cn
Shanghai Institute for Advanced Communication and Data Science,
Key laboratory of Specialty Fiber Optics and Optical Access Networks,
Shanghai University, Shanghai, 200444, China

**Abstract.** Positioning is becoming a key component in many Internet of Things (IoT) applications. The main challenges and limitations are the narrow bandwidth, low power and low cost which reduces the accuracy of the time of arrival (TOA) estimation. In this paper, we consider the positioning scenario of Narrowband IoT (NB-IoT) that can benefit from observed time difference of arrival (OTDOA). By applying the deep learning based technique, we explore the generalization and feature extraction abilities of neural networks to tackle the aforementioned challenges. As demonstrated in the numerical experiments, the proposed algorithm can be used in different inter-site distance situations and results in a 15% and 50% positioning accuracy improvement compared with Gauss-Newton method in line-of-sight (LOS) scenario and non-line-of-sight (NLOS) scenario respectively.

**Keywords:** Positioning, NB-IoT, observed time difference of arrival, deep neural network.

## 1 Introduction

Various type of smart terminals, such as self-driving cars [1], smart meters [2], and wearable devices [3], provide a paradigm shift in our daily works and lives. With the urgent requirement for the connected world, the IoT transmission has been identified as one of the most important technologies for future wireless networks, and the massive machine-type communication (mMTC) has been selected as one of the most important scenarios for the coming 5G communication system [4]. In order to provide a smooth evolution to the mMTC transmission, NB-IoT is proposed to offer initial IoT services with wide coverage, low power and spectrum consumption, and large connectivity [5] .

Among NB-IoT applications, including smart grid, safety monitoring and vending machines [6], the localization services become a fundamental feature and according to [7], more than 40% IoT connections will be related to location information by 2020. With the limit power and cost budget for NB-IoT terminals, the conventional global navigation satellite system (GNSS) based localization schemes are not available in the current NB-IoT systems, and the most common approach, according to 3GPP standard [8], relies on observing the time difference of downlink reference signals, which is often referred to as observed time difference of arrival (OTDOA).

Although the OTDOA based localization schemes have been successfully utilized in the traditional long-term evolution (LTE) systems [8], the extension to NB-IoT systems is not straight forward. For example, an iterative expectation maximization based successive interference cancellation algorithm has been proposed in [9] to jointly consider the residual frequency offset, the fading coefficients of different channel taps, and the time-of-arrival information of different cells. In [10], the estimation of phase differences for different frequency hopping reference signals has been utilized to reduce the positioning error. However, the achievable positioning accuracy of the above schemes is quite limited and the key positioning issues in the NB-IoT systems have not been completely investigated as explained below.

**Limited Resolution.** To maintain a low cost implementation of NB-IoT systems, the baseband sampling frequency has been reduced from the conventional LTE requirement (30.72 MHz) to much lower rates (1.92 MHz). Although this type of scheme greatly reduces the potential energy consumption, the resolution of observed reference signal time difference (RSTD) will be reduced as well, which causes a larger positioning error.

**Complexity.** Traditional algorithms, such as the famous Gauss-Newton method [11], require several iterations to achieve a satisfied positioning result, and the associated power consumption may not be suitable for some low power NB-IoT applications. Therefore, a low complexity and high accuracy algorithm will be desirable.

On the other hand, some deep learning schemes are proposed to solve positioning problem. [12] and [13] propose to use deep learning algorithm instead of KNN to improve the accuracy of fingerprint-based positioning. [14] uses deep learning to estimate the absolute position directly from the raw channel impulse response (CIR) data. All of these works perform well with multiple antennas. However, high precision positioning is also important in the single input single output (SISO) system.

In this paper, we propose a deep learning inspired positioning framework to address the above challenges in NB-IoT systems. Different from the traditional brute-force application of neural networks, we propose to use calculated RSTD results as commonly adopted in the OTDOA scheme rather than realtime measured signal strength to achieve better positioning performance. Meanwhile, we also exploit the generalization ability of neural networks and evaluate the positioning performance under different inter-site distances. As demonstrated in the numerical experiments, our proposed scheme can achieve as much as 15% and 67% accuracy improvement in LOS scenario and NLOS scenario respectively, if compared with traditional Gauss-Newton methods.

The rest of this paper is organized as follows. Section 2 provides some preliminary information about RSTD measure and Gauss-Newton method. Section 3 presents the deep learning based framework for position estimation, and the numerical examples are given in Section 4. Finally, we conclude this paper in Section 5.

## 2 Preliminaries

Consider an OTDOA based positioning scheme as shown in Fig. 1, where each user equipment (UE) is assumed to be connected with multiple base stations (BSs). Since the perfect synchronization between UE and BSs is in general difficult to obtain in the practical systems, RSTDs, i.e., the time-of-arrival (TOA) differences of reference signals from different BSs, are commonly adopted in the positioning process. Given the measured RSTDs, we can draw different hyperbolic curves and estimate the UE's position based on the inter section region formed by different curves as shown in Fig. 1. In this section, we briefly elaborate the composition of the RSTD estimation error and the traditional Gauss-Newton algorithm to compute the estimated positions.

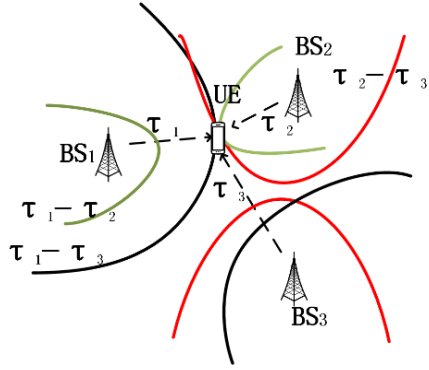

**Fig. 1.** Illustration of the OTDOA based positioning scheme. The UE can receive the signals of 3 BSs and obtain TOAs, e.g., $\tau_1$, $\tau_2$ and $\tau_3$. Then the three hyperbolic curves can be drawn by the time differences. The intersection of the three hyperbola is the position of the UE.

### 2.1 Composition of RSTD Error

The OTDOA based scheme is to estimate the two-dimensional (2D) UE position, based on the pre-known positions of BSs, e.g. $\{\boldsymbol{p}_i = [x_i, y_i]^T\}$, and the observed RSTDs, e.g. $\{\Delta\tau_i^\star\}$. RSTD is obtained by subtracting the TOA of the reference BS and the neighboring BS, e.g,

$$\Delta\tau_i^\star = \tau_i^\star - \tau_0^\star \tag{1}$$

where $\tau_i^\star$ ($i = 0,1,\cdots,N_{BS} - 1$) indicates the measured TOA, and $N_{BS}$ is the total number of observed BSs. In practical system, the RSTD is often computed by the additive noise as given below, for $i = 1,2,\cdots,N_{BS} - 1$,

$$\Delta\tau_i^\star = \frac{1}{c}\big(||\boldsymbol{p} - \boldsymbol{p}_i|| - ||\boldsymbol{p} - \boldsymbol{p}_0|| + n_{i,0} + e_i\big) \tag{2}$$
$$i = 0,1,\cdots,N_{BS} - 1$$

where $c$ is the speed of light, $n_{i,0}$ and $e_i$ are the difference of measurement error including the sampling error and NLOS error between the serving BS and the neighbouring BS respectively.

### 2.2 Gauss-Newton Method

Mathematically, the position estimation process can be obtained by minimizing the overall square errors with respect to the observed TOA results, e.g.,

$$\begin{aligned}\boldsymbol{p}^\star &= \arg\min_{\boldsymbol{p} \in \mathbb{R}^2} \sum_{i=1}^{N_{BS}-1} |c\Delta\tau_i^\star - h_i(\boldsymbol{p})|^2 \\ &= \arg\min_{\boldsymbol{p} \in \mathbb{R}^2} \sum_{i=1}^{N_{BS}-1} |u_i(\boldsymbol{p})|^2 \\ &= \arg\min_{\boldsymbol{p} \in \mathbb{R}^2} \sum_{i=1}^{N_{BS}-1} |u_i(\boldsymbol{p}) + \nabla u_i(\boldsymbol{p})(\boldsymbol{p}^\star - \boldsymbol{p})|^2 \end{aligned} \tag{3}$$

where $\boldsymbol{p}^\star = [x^\star, y^\star]^T$ denotes the estimated position of UE, and $h_i(\boldsymbol{p})$ indicates the distance differences with respect to different BSs, which is given by,

$$h_i(\boldsymbol{p}) = \sqrt{(\boldsymbol{p} - \boldsymbol{p}_i)^T(\boldsymbol{p} - \boldsymbol{p}_i)} - \sqrt{(\boldsymbol{p} - \boldsymbol{p}_0)^T(\boldsymbol{p} - \boldsymbol{p}_0)} \tag{4}$$

In the Gauss-Newton based approach [11], the estimated position in the $(k+1)^{th}$ iteration, $\boldsymbol{p}^{(k+1)}$, is obtained via the following iterative equation

$$\boldsymbol{p}^{(k+1)} = \boldsymbol{p}^{(k)} + \beta_k \left( \boldsymbol{H}^T(\boldsymbol{p}^{(k)}) \boldsymbol{H}(\boldsymbol{p}^{(k)}) \right)^{-1} \boldsymbol{H}^T(\boldsymbol{p}^{(k)}) \left( c\Delta\tau - \boldsymbol{h}(\boldsymbol{p}^{(k)}) \right) \tag{5}$$

where $\beta_k$ denotes the step size with certain convergence property, $\Delta\tau = \left[ \Delta\tau_1, \Delta\tau_2, \cdots, \Delta\tau_{N_{BS}-1} \right]^T$, and $\boldsymbol{h}(\boldsymbol{p}) = \left[ h_1(\boldsymbol{p}), h_2(\boldsymbol{p}), \cdots, h_{N_{BS}-1}(\boldsymbol{p}) \right]^T$. $\boldsymbol{H}(\boldsymbol{p})$ is the Jacobian matrix of $\boldsymbol{h}(\boldsymbol{p})$, which is defined as,

$$\boldsymbol{H}(\boldsymbol{p}) = \begin{bmatrix} \frac{\partial h_1(\boldsymbol{p})}{\partial x} & \frac{\partial h_1(\boldsymbol{p})}{\partial y} \\ \frac{\partial h_2(\boldsymbol{p})}{\partial x} & \frac{\partial h_2(\boldsymbol{p})}{\partial y} \\ \vdots & \vdots \\ \frac{\partial h_{N_{BS}-1}(\boldsymbol{p})}{\partial x} & \frac{\partial h_{N_{BS}-1}(\boldsymbol{p})}{\partial x} \end{bmatrix} \tag{6}$$

## 3 Deep learning for position estimation

The relationship between RSTD and UE location is highly nonlinear, but the Gauss-Newton algorithm is only a quadratic convergence which will result in an increase in positioning error. On the other hand, when the TOA is NLOS, the optimization algorithm for position estimation is non-convex. Thus, We try to solve these problems with the neural network which is very popular recently. Deep neural networks (DNN) have good capability of classification and regression. DNN's classification ability has been well shown in fingerprint-based positioning [15]. In our algorithm, we use DNN's great regression ability to fit the relationship between RSTD and UE's position.

### 3.1 TOA Error Distribution

In LTE system, the sampling rate is $F_s = 30.72\text{MHz}$ and each sampling period is $T_s = 1/F_s$. However, in NB-IoT system the sampling-rate could be defined as $F_s = 1.92\text{MHz}$ in order to save battery-power. We consider scenarios without inter-cell interference. We did 1000 Monte Carlo simulations, and recorded the TOAs of 7 BSs for each simulation. The error distribution of TOA is shown in Table. 1.

What we can learn from the Table. 1 is that the TOA estimation errors are mainly caused by low sampling rate and NLOS. In practice, however, higher sampling rates mean more power consumption, which is not recommended in IoT devices, and if we can mitigate the effect of NLOS, we can greatly improve the accuracy of positioning. We define the NLOS scenario when the TOA error is greater than 0 Ts. In OFDM systems, some NLOS identification technologies [16, 17] have already performed well. In this article we assume that the recognition result of NLOS is prior information, and let LOS status $s = 0, 1$ where 1 indicates NLOS and 0 indicates LOS.

**Table 1.** TOA error distribution under different kinds of channel

| TOA Error | $-1T_s$ | $0T_s$ | $1T_s$ | $2T_s$ |
|---|---|---|---|---|
| AWGN | 90 | 6842 | 68 | 0 |
| EPA | 121 | 6737 | 138 | 4 |
| EVA | 376 | 3586 | 2502 | 559 |

**3.2 Data Collection and Preprocessing**

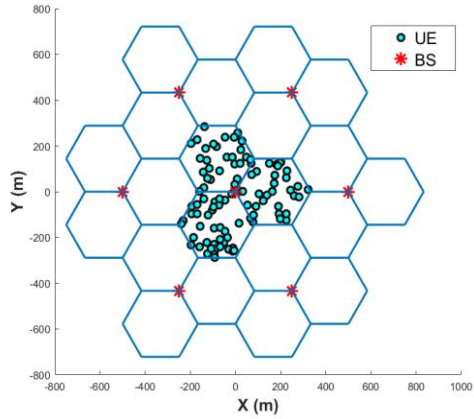

**Fig. 2.** The topology of the simulation. There are a total of 7 BSs, and the intermediate BS is regarded as a serving BS. We randomly generate UEs around the serving BS.

We can regard DNN as a black box f(·) through a lot of data training. In the problem of OTDOA positioning, the function can be written as $\boldsymbol{p} = f(\cdot)$.

In the first step, we need a lot of data sets about W and $\boldsymbol{p}$ mapping on f(·) to train the network. We consider the case of $N_{BS}$, that is, the number of BSs is 7. As shown in Fig. 2, UEs are randomly generated in cells around the intermediate BS. In order to generalize different inter-site distance situations, we normalize the input so that the normalized input parameter can be written as,

$$\boldsymbol{w} = \frac{[c\Delta\tau_1^\star, x_1, y_1, \cdots c\Delta\tau_6^\star, x_6, y_6, D_{cell}s_1, \cdots D_{cell}s_6]}{D_{cell}} \quad (7)$$

where $D_{cell}$ represents the distance of two adjacent cells, and $s_i$ indicates the $i^{th}$ BS's LOS status. It is easy to find that the input parameters of BS's coordinates are certain value after normalization, but we still use these parameters as input parameters because they contain the position information even if it is not necessary. We generate 100,000 sets of $\boldsymbol{w}$ and their corresponding coordinates to train the network.

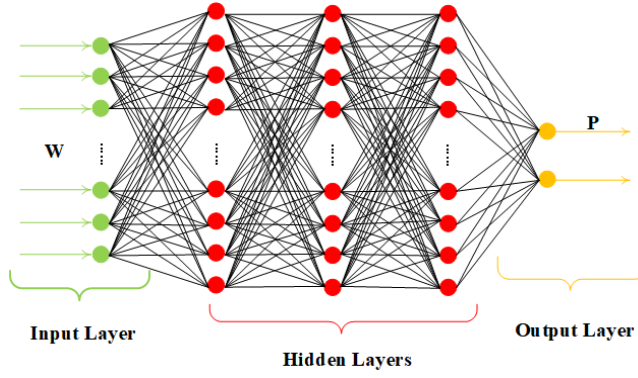

**Fig. 3.** Schematic diagram of the DNN structure. The input parameter $w$ contains RSTD and BS position information, and the output $p$ represents the estimated position.

### 3.3 Neural Network Configuration

**Table 2.** Network structure

| Layers | Output Shape | Activation |
|---|---|---|
| Input layer | 18 | - |
| Fully connected layer 1 | 18 | Relu |
| Fully connected layer 2 | 32 | Relu |
| Fully connected layer 3 | 16 | Relu |
| Fully connected layer 4 | 8 | Relu |
| Output layer | 2 | Linear |

The network diagram is shown in Fig. 3. and the network structure of the DNN in our experiments is shown in Table 2. An input layer, an output layer and 4 hidden layers form this network, and these four hidden layers have 18, 32, 64, and 8 neural units respectively. The target is to find a function f(·) and to make the estimated coordinates and real coordinates as close as possible. The optimization function can be written as,

$$\boldsymbol{p}^\star = \arg \min_{\boldsymbol{p}^\star \in \mathbb{R}^2} |\boldsymbol{p} - \boldsymbol{p}^\star|^2 \tag{8}$$

Thus, we choose mean square error (MSE) as loss function and the loss function can be given by,

$$\mathcal{L} = \frac{1}{N_{batch}} \arg \min_{\boldsymbol{p}^\star \in \mathbb{R}^2} \sum_{m=1}^{N_{batch}} \left|\boldsymbol{p}^{(m)} - \boldsymbol{p}^{\star(m)}\right|^2 \tag{9}$$

where $N_{batch} = 32$ is the batch size for each training step in our experiments, $\boldsymbol{p}^{(m)}$ and $\boldsymbol{p}^{\star(m)}$ are the $m^{\text{th}}$ training position data. In our experiment, We only use 6 layers. Thus, the training speed and calculating speed are both faster than the Gauss-Newton algorithm. Use the generated data sets to train the network, and the trained network can be used to estimate UE's position.

### 3.4 Model Generalization

Because the BSs' coordinates are known values, the normalized distance differences are the main parameters affecting network training results and positioning accuracy. It can be proved that the actual distance difference between the serving BS and the neighbouring BS to the UE is in the range of $[-D_{cell}, D_{cell}/3]$. The time difference is converted to RSTD after sampling, and its range of values can be expressed as,

$$\Delta\tau_i^\star \in \frac{\left\{<-\frac{F_s D_{cell}}{c}>, <-\frac{F_s D_{cell}}{c}>+1, \cdots <\frac{F_s D_{cell}}{3c}>\right\}}{F_s} \tag{10}$$

where $<\cdot>$ indicates rounding. After normalization, the range of distance difference can be written as,

$$\frac{c\Delta\tau_i^\star}{D_{cell}} \in \frac{\left\{<-\frac{F_s D_{cell}}{c}>, <-\frac{F_s D_{cell}}{c}>+1, \cdots <\frac{F_s D_{cell}}{3c}>\right\} \cdot c}{F_s D_{cell}} \tag{11}$$

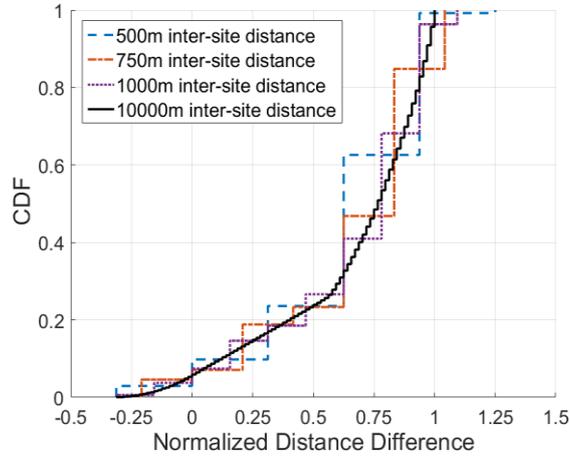

**Fig. 4.** The CDF of normalized distance difference $\frac{c\Delta\tau_i^\star}{D_{cell}}$.

After the distance difference is normalized, it is distributed in a similar range, and in the case of large inter-site distance, the distribution contains more information. We have carried out experiments when the inter-site distance is 500m, 750m, 1000m, and 10000m. The cumulative distribution function (CDF) of the normalized distance difference $\frac{c\Delta\tau_i^\star}{D_{cell}}$ is shown in Fig. 4. It is like sampling the input space of a neural network. Moreover, the large distance model can carry out more fine-grained sampling. A more fine-grained sampling of input means that the DNN can fit a more precise function f(·) after training. Therefore, we use 10000m inter-site distance normalized data sets to train the network, and evaluation of the performance of this network by 500m, 750m, 1000m inter-site distance data sets.

## 4 Numerical Results

In this section, we provide numerical results to compare the performance of the DNN algorithm with the Gauss-Newton one. Gauss-Newton's algorithm. The simulation parameters are summarized in Table 3.

**Table 3.** Scenario parameters

| Layers | Output Shape |
|---|---|
| Network synchronization | Perfect synchronization |
| Cyclic prefix | Normal |
| Number of BS/device antennas | 1/1 |
| Macro transmit power | 46 dBm for 1.92 MHz |
| Terminal noise density | -174 dBm/Hz |
| Number of PRS | 1 |
| Consecutive PRS subframes | 1 |
| PRS muting | Yes |

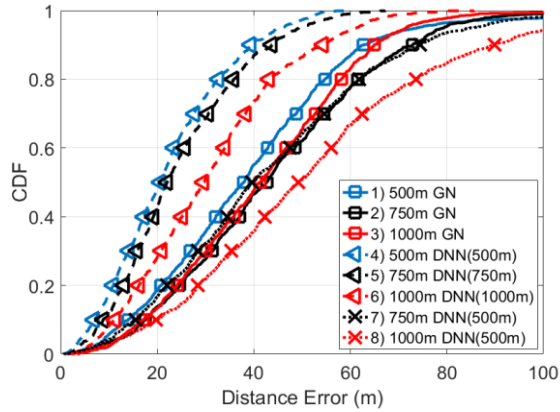

**Fig. 5.** The performance comparison between two algorithms when the inter-site distance is different. The results come from Monte Carlo simulation of 1000 random UE locations. 1), 2), 3) represent the performance of Gauss-Newton algorithm in three different inter-site distance cases. 4), 5), 6) represent the performance of the DNN algorithm which are trained and tested by the same inter-site distance data sets. 7) and 8) represent the performance of the DNN algorithm which are trained by 500m inter-site distance data sets but tested by 750m or 1000m data sets.

## 4.1 Results of the DNN Algorithm

Fig. 5 illustrates the CDF of the horizontal positioning accuracy of two different algorithms. We consider three different kinds of distance and the performance of the DNN algorithm is better than the performance of Gauss-Newton algorithm when we use the same distance data sets to train and test the network. Lines 1 to 6 indicate that the positioning accuracy of the DNN algorithm can achieve about 25% accuracy improvement.

However, when the inter-site distance of the training and test data sets is different, the performance of the DNN algorithm is very poor even if the data sets are normalized. For example, in the experiment of 750m inter-site distance, the average positioning accuracy is 24.5m when the network is trained by 750m inter-site distance data sets and it is only 43.2m when the network is trained by 500m inter-site distance data sets. That is because when we use the data sets of 500m inter-site distance, the data sets are not general enough to allow the network to learn the characteristics of the inter-site distance of 750m and 1000m.

In order to evaluate the positioning ability of the DNN algorithm under different number of BSs, we set the number of BSs to 4, 5, and 6 respectively. The performance of the proposed algorithm is shown in Fig. 6. It indicates that the more BSs participating in the positioning, the higher the positioning accuracy will be and the proposed the DNN algorithm performs better than the Gauss-Newton algorithm.

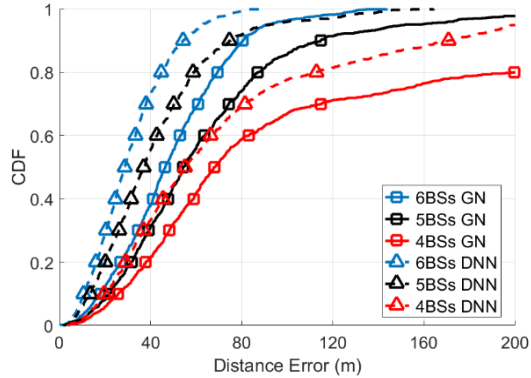

**Fig. 6.** The performance comparison between the Gauss-Newton algorithm and the DNN algorithm when the number of BSs participating in positioning is different.

## 4.2 Results of Generalized Networks

After that, we use 10000 m inter-site distance data sets to train the DNN, and use 500m, 750m, 1000m distance data sets to test the network. The result is shown in Fig. 7. In this case, the positioning accuracy of DNN is improved by about 15%. It achieves significant gains under different inter-site distances. Therefore, this method successfully generalizes the network at the cost of some positioning accuracy, enabling high accuracy positioning on the same network even if the inter-site distance is different.

We can also find a trade-off relationship between the inter-site distance of the training sets and the positioning accuracy of the test sets. It indicates that the generalization ability of the network should be at the expense of positioning accuracy and if we want to cover more inter-site distance situation, the positioning accuracy will be reduced.

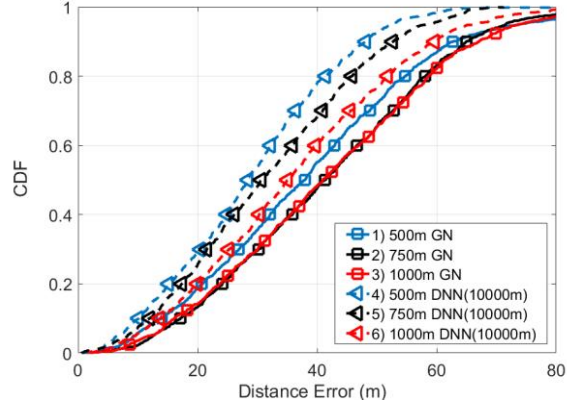

**Fig. 7.** The performance comparison between the Gauss-Newton algorithm and the DNN algorithm which is trained by 10000m inter-site distance data sets. 1), 2, 3) represent the performance of Gauss-Newton algorithm in three different inter-site distance cases. 4), 5), 6) represent the performance of the DNN algorithm which are trained by 10000m inter-site distance data sets but tested by 500m, 750m and 1000m data sets.

### 4.3 Results of NLOS scenario

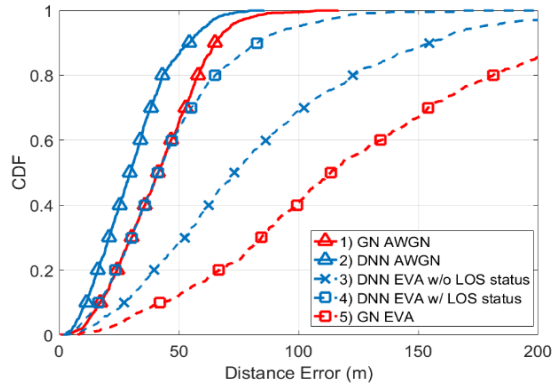

**Fig. 8.** The performance comparison between the Gauss-Newton algorithm and the DNN algorithm under EVA channel. The positioning accuracy will decrease because of the NLOS scenario. Line 4 indicates when the DNN input contains the LOS status $s$, the positioning accuracy can be greatly improved.

We use the AWGN channel to represent the LOS scenario, and the EVA channel to represent the scenario where LOS and NLOS coexist. Fig. 8 illustrates the CDF of the horizontal positioning accuracy of two different algorithms under AWGN and EVA channels. Line 5 indicates that the positioning accuracy will decrease because of the NLOS scenario. However, when the input of DNN contains the LOS status $s$, the positioning accuracy can be greatly improved by about 67% and 50%, if compared with Gauss-Newton algorithm and DNN algorithm without LOS status input, respectively. This figure shows that the LOS status can be beneficial to improve the positioning accuracy.

## 5 Conclusion

In this paper, we propose to use DNN for location estimation when the TOA estimation accuracy can not be improved in 1.92 MHz sampling rate. In order to solve the DNN generalization problem, we analyze the CDF of the distance difference, and use the large inter-site distance data sets to train the network. We also use the LOS status $s$ as input to improve the positioning accuracy in the NLOS scenario. The proposed algorithm can improve positioning accuracy by about 15% in LOS scenario and 67% in NLOS scenario compared with Gauss-Newton method.

**Acknowledgement**

This work was supported by the National Natural Science Foundation of China (NSFC) Grants under No. 61671011, 61701293 and No. 61871262, 61401266, the National Science and Technology Major Project Grants under No. 2018ZX03001009, the Huawei Innovation Research Program (HIRP), and research funds from Shanghai Institute for Advanced Communication and Data Science (SICS).